\documentclass[12pt]{article}
\usepackage[explicit]{titlesec}
\usepackage{etoolbox}
\usepackage{multicol}
\usepackage{hyperref}
\usepackage[natbibapa, nosectionbib]{apacite} 

\usepackage{lmodern}
\usepackage{fancyvrb}
\usepackage{amssymb,amsmath}

\usepackage[a4paper, portrait, margin=25mm]{geometry}
\usepackage{ragged2e}
\usepackage{cprotect}
\usepackage[dvipsnames]{xcolor}

\usepackage{tabularx}
\usepackage{booktabs}
\usepackage{microtype}

\usepackage[symbol]{footmisc}

\makeatletter
\patchcmd{\@maketitle}{\LARGE}{\Huge}{\typeout{OK 1}}{\typeout{Failed 1}}
\patchcmd{\@maketitle}{\large \lineskip}{\Large \lineskip}{\typeout{OK 2}}{\typeout{Failed 2}}
\makeatother

\makeatletter
\newcommand{\verbatimfont}[1]{\renewcommand{\verbatim@font}{\ttfamily#1}}
\makeatother

\usepackage{bigfoot}

\titleformat{name=\paragraph, numberless}{}{\Large\normalfont\sffamily #1}{.5em}{}

\titleformat{name=\section}{}{\vspace{.8ex} \Large\normalfont\sffamily \thesection \hspace{0.2em} #1 }{.5em}{}
\titleformat{name=\subsection}{}{\vspace{.8ex} \large\normalfont\sffamily \thesubsection \hspace{0.2em} #1 }{.5em}{}

\newcommand{\newAuthor}[3]{
\noindent
#1 \\
#2 \\
#3 \\
}

\title{\textbf{\uppercase{The match file format: \\ Encoding Alignments \\ between Scores and \\ Performances}}}

\usepackage{xspace}
\usepackage{listings}

\def\ie{\textit{i.e.,}\xspace}
\def\eg{\textit{e.g.,}\xspace}

\def\etc{\textit{etc}\xspace}

\def\MEI{\textsf{MEI}\xspace}
\def\Kern{\textsf{Humdrum **kern}\xspace}

\def\MIDI{\textsf{MIDI}\xspace}
\def\MusicXML{\textsf{MusicXML}\xspace}
\def\Music21{\textsf{Music21}\xspace}

\def\Partitura{\textsf{Partitura}\xspace}

\date{}

\usepackage{graphicx}
\graphicspath{{images/}}

\begin{document}

\maketitle

\begin{multicols}{3}
    
    \newAuthor{Francesco Foscarin\footnote{equal contribution.}}{Johannes Kepler University}{francesco.foscarin@jku.at}    
    
    \vspace{1cm}
    
    \newAuthor{Carlos Cancino-Chac\'{o}n\footnotemark[1]}{Johannes Kepler University}{carlos\_eduardo.\\cancino\_chacon@jku.at}
    
    \columnbreak
    
    \newAuthor{Emmanouil Karystinaios\footnotemark[1]}{Johannes Kepler University}{emmanouil.karystinaios@jku.at}
    
    \vspace{1cm}
    
    \newAuthor{Maarten Grachten}{Independent Researcher}{maarten.grachten@gmail.com}    
    
    \columnbreak
    
    \newAuthor{Silvan David Peter\footnotemark[1]}{Johannes Kepler University}{silvan.peter@jku.at}
    
    \vspace{1cm}
    
    \newAuthor{Gerhard Widmer}{Johannes Kepler University}{gerhard.widmer@jku.at}    
\end{multicols}





\paragraph{Abstract}

This paper presents the specifications of \textit{match}: a file format that extends a MIDI human performance
with note-, beat-, and downbeat-level alignments to a corresponding musical score. 
This enables advanced analyses of the performance that are relevant for various tasks, such as expressive performance modeling, score following, music transcription, and performer classification.
The match file includes a set of score-related descriptors that makes it usable also as a bare-bones score representation.
For applications that require the use of structural score elements (\eg voices, parts, beams, slurs), the match file can be easily combined with the symbolic score. 
To support the practical application of our work, we release a corrected and upgraded version of the Vienna4x22 dataset of scores and performances aligned with match files.



\renewcommand{\thefootnote}{\arabic{footnote}}
\setcounter{footnote}{0}

\paragraph{Introduction}

In this work, we present \textit{match}: a file format for a complete and robust encoding of symbolic music alignments.
The term \textit{symbolic} refers to the class of musical data types that explicitly represent a set of elements from common music notation. 
Such a set must include at least note elements with their temporal position and (where applicable\footnote{Percussive notes may not have a pitch or duration.}) pitch and duration.
Common data types that are symbolically encoded are musical scores (in \MEI, \MusicXML, \Kern, and \MIDI\footnote{Though \MIDI can only encode a partial set of the score information.} formats) and performances (in \MIDI format).
This is opposed to other data types, such as audio and raster score images, which only represent low-level 
information, such as amplitude over time, and pixel RGB values, respectively.
Several works in music information retrieval (MIR)  make use of symbolic data types
as we can expect a more explicit representation of music to produce more efficient and musically interpretable systems.

A symbolically encoded performance, in short, \textit{symbolic performance}, consists of a sequential representation of notes  with a position and duration given in terms of real physical time. It is produced by recording a human performance with musical instruments fitted with proper sensors, such as MIDI keyboards, MIDI drums, Disklavier grand piano, and, to some extent, guitars with MIDI pickups. Those instruments can capture and explicitly represent the time and dynamic deviations that are natural aspects of an expressive performance~\citep{honing2001time}.
On the other side, a \textit{symbolic musical score} expresses note positions in \textit{musical units} such as fractions of quarter notes and beats
and arranges them in temporal and organizational structures such as measures, beats, sub-beats, parts, and voices. 
It also explicitly represents dynamics and temporal directives and other high-level musical features such as time signature, pitch spelling, and key signatures.

\begin{figure}[t]
\includegraphics[width=0.90\textwidth]{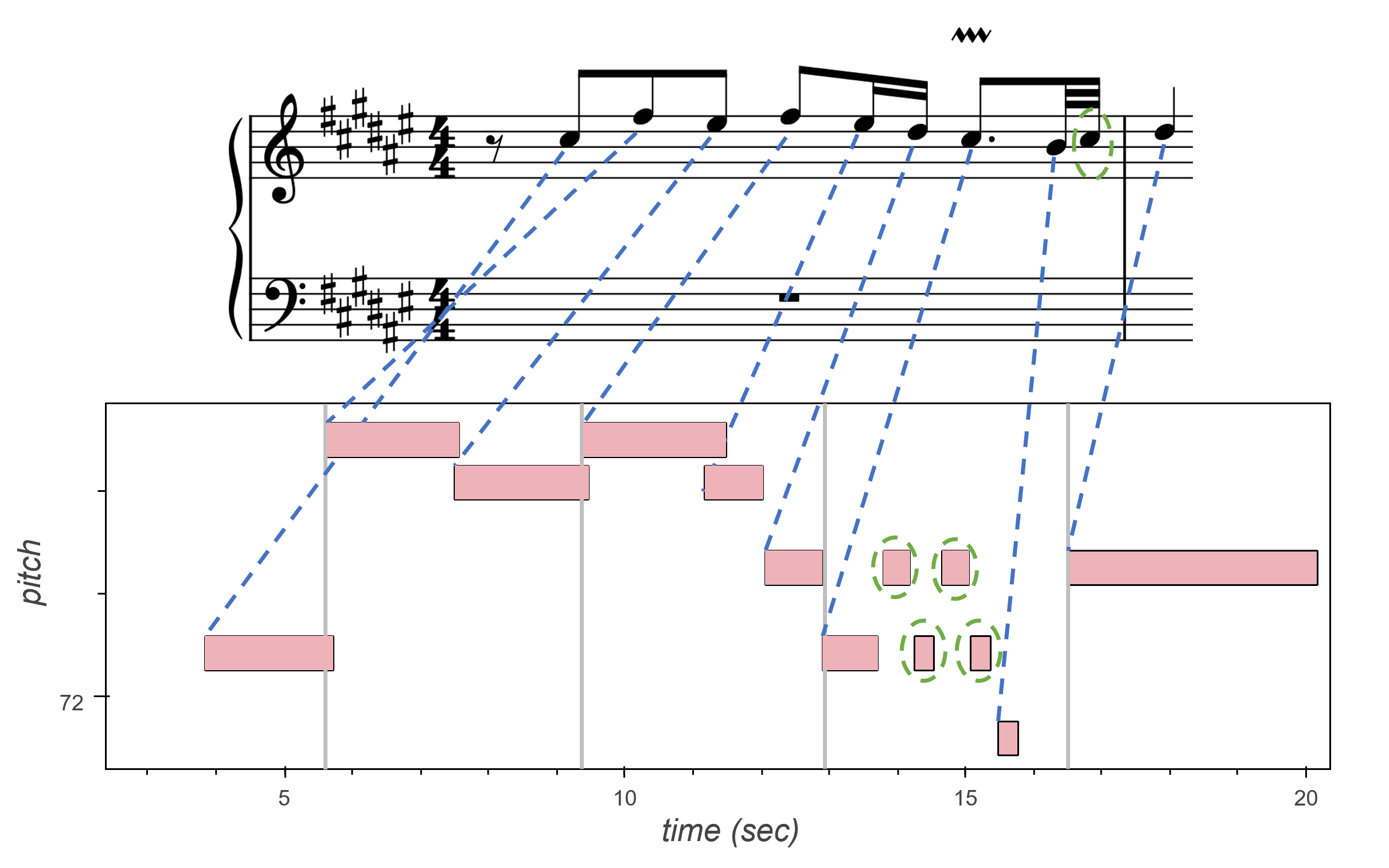}
\centering
\caption{An example note-level alignment between a score and a performance of a score with a real piano performance. From Bach Fugue 13 in F sharp major, BWV 858. Aligned notes are connected blue dotted lines, while notes that are only in the score (deletion) or in the performance (insertion) are circled in green. Beat positions are marked with grey vertical lines in the performance.}
\label{fig:initial_example}
\end{figure}

The structured representation of the score and the expressive nuances encoded in the performance are complementary elements that can be combined to enable advanced musical analysis. Typical MIR tasks that benefit from it are expressive performance modeling, score following, music transcription, and performer classification.
However, to fully leverage this information, we need an alignment between each corresponding element in a score and a performance (see Figure~\ref{fig:initial_example} for a short example). Some fully automated techniques have been proposed to produce alignments at note-level~\citep{nakamura2017performance, gingras:2011,Chen:2014}, but, as they struggle in certain situations, it is still common to go through a manual annotation/correction phase performed by experts~\citep{asap-dataset}.

In this paper, we assume a score-performance alignment is given, for example as a result of a manual or semi-automatic alignment, and we address the problem of encoding it in a format that focuses on completeness, usability, and robustness. We name this format \textit{match}. 
Completeness is ensured by explicit handling of repetitions structures and by including in the match file the entire list of notes in the performance, even if they are not present in the score, for example as a result of embellishments, player mistakes, or incomplete scores.
For usability, we reduce the technical difficulties of operating across multiple files, by encoding in the match file a lossless representation of the performance, enhanced with a bare-bones representation of the score. This enables the usage of match files as a stand-alone representation for all tasks that focus on pitch and duration information.
Furthermore, the extra score information acts as a redundancy safety layer to improve the robustness of the alignments for applications that need to link to the symbolic score to use other
score elements (e.g., voices, parts, beams, slurs). 
On the contrary, a more naive encoding that only points at score positions with beats and measures number would fail in case of minor score modifications such as the splitting or time modification of a measure. 

The Match format is text-based, sequentially structured, and human understandable. This enables the visual inspection and manual editing of its content, and eases its integration in different applications. For Python-based research, the usage of match files is further simplified by 
the \Partitura package~\citep{partitura_late_demo} that offers off-the-shelf parsing and processing of this format.
To support the practical application of our work, we release a corrected and upgraded version of the Vienna 4x22 dataset of symbolic performances and aligned scores~\citep{vienna4x22}.

The remainder of this paper is organized as follows: in Section~\ref{sec:related_work} we compare with other relevant research and highlight how our system can solve typical problems in this field. In Section~\ref{sec:match_file} we detail the match file format and Section~\ref{sec:dataset} we present the dataset and match file parsing with Partitura. Finally, in Section~\ref{sec:conclusions} we draw some conclusions and discuss future work.

\section{Related work}\label{sec:related_work}

Some datasets exist that contain alignments between scores and performances.
For example, the expressive performance dataset~\citep{Marchini:2014aa} contains note-level alignments (onset, offset and pitch), the Mazurka dataset~\citep{cook2007performance} provides beat-level alignments and the ASAP dataset \citep{asap-dataset} contains alignments at the measure level. To store them, these datasets use text files (one file for each monophonic part),  Excel spreadsheets, and JSON files, respectively. None of those formats is directly extensible to include a complete set of alignments. The absence of a unique encoding forces researchers to spend time learning new formats and writing new code every time they need to target a different dataset.

A proposal for a general and easy-to-use encoding of note alignments is made by \citet{devaney2019encoding}, by extending the musical score (in \MEI or \Kern format) with some performance-related information.
A problem that arises from this approach is about encoding efficiency: if we are considering multiple performances of the same score, we would need to make multiple duplicates of the entire musical score. This creates a lot of repeated information, increases memory usage, and produces consistency problems if we want, for example, to correct a notation error in the score. Another problem is that this format can't encode information on performed notes that are not on the score, \eg embellishments or player mistakes.
Similarly to this approach, our match file contains enough information on the score and performance to be used as a stand-alone representation for many tasks. However, instead of extending the score, we work in the opposite direction by extending the performance with a bare-bones score representation.

\cite{nakamura2017performance} present a system that is able to automatically align two MIDI performances or a \MusicXML score and a performance. The output of such a system is a list of references, similar to the main part of our match file. Differently from our approach, however, Nakamura uses real numbers with (6 decimal digits precision) to identify the onset and offset of notes in the MIDI file. This limited precision can lead to rounding problems and force the usage of approximate equality functions to retrieve the corresponding note in the MIDI. Instead, our approach ensures a lossless encoding of performance time positions with MIDI ticks that are already used in the MIDI file.



Older versions of the match file have been around for some time\footnote{\url{http://dx.doi.org/
10.21939/4X22 }} but no official reference or documentation is available. The initial format was created to encode manually annotated note alignments between  pieces performed on computer-monitored pianos such as Boesendorfer SE 290 and their corresponding scores. It was developed in \textit{prolog}, a popular language at the time for database creation and retrieval of information. For this reason, each event or entry in the match file format is represented on a new line with a dot determining the end of a prolog-like compound term.
In this paper, we propose an updated version of the match file (version 1.0.0) that includes support for repetitions and time point alignments (\eg beats and downbeats) and solves other small issues that were found in practical usage. This paper has among its goals to be a formal reference to this encoding format for a more widespread utilization in the research community.


\section{Match files}\label{sec:match_file}

Match is a text-based file format developed to encode a complete and robust alignment between a symbolic performance and a corresponding musical score.
We start the description of the format with a high-level perspective on how it handles alignments and repetition structures. Then we detail how the information is encoded.



\subsection{Note and time alignment}

Match files contain alignments at two levels: notes and time points (\eg beat and measure).

For note-level alignments, the match file encodes a mapping $match$ between the notes in a performance and the corresponding one in a score.
Formally, let us consider the sets $P$ and $S$ of all notes in a symbolic performance and the corresponding musical score.
$match$ is a partial function over $P$ and $S$. 
For a performance that aligns perfectly with a score, $match$ becomes a total function, \ie it is defined $\forall e \in P$. 
However, due to player mistakes, embellishments, or incomplete scores, it may happen that some events in $P$ do not have a corresponding event in $S$. 
The partial function is non-surjective, as there are multiple events in the score (\eg time signature annotations, barlines, etc.) that are not in the performance and there is the possibility of omitted notes by the performer.
Moreover, $match$ can be non-injective if there are repetitions, where multiple events in the performance map to the same element in the score.

The case of time alignments is different because both score and performance time are continuous domains and a function between them cannot be easily annotated by music experts. Time level alignments can be seen as a set of samples from this function. 
Creators of the match files are free to select the granularity level they want to encode, usually at the downbeat or beat level.

\subsection{Repetition structures}
Handling repetitions for a performance and a corresponding score is a complex task due to the arbitrary way they can be interpreted. For instance, a performer may play the entire ``unfolded'' piece, while another may skip some repetitions.
Other works, \eg \citet{asap-dataset}, create different score versions by manually removing the repetition marks that are not played in the performance. However, this approach produces multiple scores of the same piece, which complicates the comparison of the related performances.

By using match files, we do not modify musical scores; instead, we encode in the match file an unfolded (or reduced) score that matches the aligned performance. This is paired with an explicit representation of repetitions based on \textit{sections}.\footnote{Sections in match files have a similar role to \MEI \verb|<section>| (and \verb|<ending>|) elements; however, they cannot be nested, and their usage is limited to repetition structures.} A score is segmented in multiple sections by repetition signs such as left repeat, right repeat, volta start, volta end, and navigation directions such as al Coda, dal Segno, da Capo, \etc. This simplifies the task of comparing performances of the same piece with different repetition structures.

\begin{figure}
    \noindent
    \begin{minipage}{.5\linewidth}
    \begin{Verbatim}[commandchars=\\\{\}, fontsize=\scriptsize]
    \colorbox{Lavender}{info(matchFileVersion, 1.0.0).}
    info(composer,Bach)
    info(piece,Fugue 13 BWV858)
    info(midiClockUnits,480).
    info(midiClockRate,500000).
    \colorbox{yellow}{scoreprop(timeSignature,4/4,1,0.0).}
    scoreprop(keySignature,F# Maj,1,0.0).
    \colorbox{CornflowerBlue}{snote(n0,[C,#],5,1:1,1/8,1/8,0.5,1.0,[])-note(0,73,1104,1647,43,0,0).}
    \colorbox{lightgray}{stime(1:2,0,1,beat)-ptime([1620]).}
    snote(n1,[F,#],5,1:2,0,1/8,1.0,1.5,[])-note(1,78,1620,2180,51,0,0).
    snote(n2,[E,#],5,1:2,1/8,1/8,1.5,2.0,[])-note(2,77,2160,2727,56,0,0).
    stime(1:3,0,2,beat)-ptime([2704]).
    snote(n3,[F,#],5,1:3,0,1/8,2.0,2.5,[])-note(3,78,2704,3308,55,0,0).
    snote(n4,[E,#],5,1:3,1/8,1/16,2.5,2.75,[])-note(4,77,5,3217,3464,56,0,0).
    snote(n5,[D,#],5,1:3,3/16,1/16,2.75,3.0,[])-note(5,75,5,3472,3716,55,0,0).
    stime(1:4,0,3,beat)-ptime(3716).
    snote(n7,[C,#],5,1:4,0,3/16,3.0,3.75,[])-note(6,73,5,3716,3949,58,0,0).
    \colorbox{LimeGreen}{insertion-note(7,75,3972,4084,58,0,0).}
    insertion-note(8,74,41024186,61,0,0).
    insertion-note(9,75,4221,4335,54,0,0).
    insertion-note(10,73,4341,4425,4425,63,0,0).
    snote(n8,[B,n],4,1:4,3/16,1/32,3.75,3.875,[])-note(11,71,4456,4542,55,0,0).
    \colorbox{LimeGreen}{snote(n9,[C,#],5,1:4,7/32,1/32,3.875,4.0,[])-deletion.}
    stime(2:1,0,4,downbeat)-ptime(4752)
    snote(n17,[D,#],5,2:1,0,1/4,4.0,5.0,[])-note(13,75,4752,5808,55,0,0).
    \colorbox{Red}{section(0.0,4.0,0.0,4.0,[])}
    \colorbox{YellowOrange}{sustain(17140,31,0,0).}
    sustain(17160,49,0,0).
    \end{Verbatim}
    \end{minipage}
    \cprotect\caption{An example of match file corresponding to the alignments of Figure    ~\ref{fig:initial_example}. Examples of different types of lines are color-coded for an easier understanding.
    
    }
    \label{fig:match_example}
\end{figure}

\subsection{File encoding}
A match file consists of a sequence of lines, each ending with a dot, and there are  five different types of lines. Figure~\ref{fig:match_example} highlights examples with different colors.
Global information lines (in pink) encode elements that are constant throughout the entire piece, for example, composer, performer, title, version, \etc. The score property lines (in yellow) contain score elements that can change, such as key signatures, time signatures, and performance directives.
Alignment lines can have two formats, \verb|snote(*)-note(*)| (in blue) and \verb|stime(*)-ptime(*)| (in grey), for note and time point alignments, respectively.
Non-aligned notes, \ie notes only in the performance or only in the score, are encoded with the keywords insertion and deletion (in green). 
Furthermore, specific alignment lines are designated for ornaments and trills, where multiple performed notes can refer to the same score notes or even a score marking.
Repetition section lines (in red) relate the unfolded score times to the original score times, and sustain pedal lines (in orange) encode pedal information .






Table~\ref{tab:align_specifications} contains value specifications for the note alignment line:

\begin{minipage}{0.1\linewidth}
    \noindent
    \begin{Verbatim}[commandchars=\\\{\}, fontsize=\scriptsize]
   
    snote(Anchor,[NoteName,Mod],Octave,Measure:Beat, Offset,Dur,OnsetInBeats,OffsetInBeats,ScoreAttrList)-
                                                                
    note(ID,MIDIpitch,Onset,Offset,Velocity, MIDIchannel, MIDItrack).
    \end{Verbatim}
\end{minipage}

The \verb|Anchor| string corresponds to existing identifiers if there are any in the musical score (\eg \MEI note ids) or generated ones. A suffix can be added in case of repetitions; for example, a score note "n23" in a repeated section, will be referred to as "n23-1" the first time and as "n23-2" the second.
The full specifications of the other line types are available at \url{https://cpjku.github.io/matchfile}.

\begin{table}
    \begin{center}
    \begin{tabularx}{\textwidth}{ccl}
    \toprule
        Value name & Value type & Description \\ 
    \midrule
        Anchor & string & Note identifier (suffix after “-” for repetition) \\ 
        NoteName & string &  Pitch class name in [C, D, E, F, G, A, B]  \\ 
        Modifier & string &  Pitch modifier in [“”, n, b, \#, bb, x]  \\ 
        Octave & integer &  Octave number in scientific pitch notation  \\ 
        Measure & integer &  Measure number (starting at 1, 0 for anacrusis)  \\ 
        Beat & integer & Integer beat number of note onset (starting at 1) \\ 
        Offset & fraction (int/int) & Offset from beat position (fraction of whole note)
        \\ 
        OnsetInBeats & float &  Onset position in contiguous beats 
        \\ 
        DurationInBeats & float & Duration in beats \\ 
        ScoreAttributesList & string &  Note attributes (“grace”, “appoggiatura”, etc.)  \\ 
        ID & integer & Note identifier \\ 
        MIDIpitch & integer & Pitch 0-127 \\ 
        Onset & integer & Time in MIDI ticks of the note on message \\ 
        Offset & integer & Time in MIDI ticks of the note off message \\ 
        Velocity & integer & Note on velocity 0-127 \\
        Channel & integer & MIDI channel 0-15 \\ 
        Track & integer & MIDI track \\
    \bottomrule
    \end{tabularx}
    \end{center}
    \caption{Values specifications for the alignment line in a match file. }
    \label{tab:align_specifications}
\end{table}

\section{Practical applications}\label{sec:dataset}
To support the usage of match files in practical applications, we present a way of parsing and processing those files in Python and an updated and corrected match file dataset.


\subsection{Handling Match files}
\Partitura~\citep{partitura_late_demo,partitura_mec}, an open-source Python package, offers off-the-shelf parsing and processing of match files.
\Partitura creates dedicated Python objects that give easy access to the information encoded in the match file, as well as lists of dictionaries relating performance and score note identifiers.
If external scores are required, \Partitura can load \MEI, \MusicXML, and \Kern scores and link notes and temporal positions by using the alignments in the match file.
\Partitura objects can be easily converted to note array or pianoroll representations. 

\subsection{Dataset}

The Vienna 4x22 was originally compiled by~\cite{vienna4x22}  and consists of 4 excerpts of solo piano pieces, each performed by 22 pianists. 
The pieces are 21 bars of Chopin Opus 10, 45 bars of Chopin Opus 38, 36 bars (the exposition) of Mozart KV331, and the full 32 bars of Schubert D783 (opus 33 No. 15). 
All performances were recorded on B\"{o}sendorfer 290 SE Grand Piano as MIDI-like data and subsequently each played note matched to its respective score note.
We release an updated and corrected version of this dataset encoded in the current version of the match format at \url{https://github.com/CPJKU/vienna4x22}.

\section{Conclusion and Future Work}\label{sec:conclusions}

In this paper, we proposed the match format for a complete and robust  encoding of the alignments between symbolically encoded musical scores and performances. Note-level alignments and time-level alignments (\eg beat and measure) are supported. Match files can be used as a stand-alone representation of score and performance to reduce the technical difficulties of operating across multiple files.
A Python package that can read and write match files is available.\footnote{\url{https://partitura.readthedocs.io/}} We also release an updated and corrected version of the Vienna4x22 dataset that contains scores and performances aligned with match files.

The match file format is actively under development. This paper marks the release of the stable version 1.0.0 but further modifications are to be expected to support more features and solve problems that arise from its usage in practical applications.  Other future works involve a graphical utility for the visualization and modification of alignments and tools to create them automatically (at least partially) given a corresponding score and performance.








\paragraph{Acknowledgements}
This project receives funding from the European Research Council (ERC) under the European Union's Horizon 2020 research and innovation programme, grant agreement No 101019375 (\textit{Whither Music?}).


\paragraph{References}

\bibliographystyle{apacite}
\bibliography{biblio}

\end{document}